\documentclass[epj]{svjour}

\usepackage{graphicx}
\usepackage{fancyhdr}
\def\makeheadbox{{%  remove EPJC header
 }}
\def\mytitle{My title} 
\def\myauthors{My name}  
\def\mytype{My type of session}
\def\mysession{My session}
\def\mytitle{Isospin asymmetry in $b \rightarrow s \gamma$} %Put your title here!
\def\myauthors{F. Mahmoudi}    %Put your name here!
\def\mytype{Contributed Talk}    
\def\mysession{Colliders - SUSY Phenomenology}
\begin{document}
\title{Supersymmetric parameter constraints from isospin asymmetry in $b \rightarrow s \gamma$ transitions}
% \subtitle{Do you have a subtitle?\\ If so, write it here}
\author{F. Mahmoudi\inst{}
% \thanks is optional - remove next line if not needed
\thanks{\emph{Email:} nazila.mahmoudi@tsl.uu.se}%
%  \and
%  Second author\inst{2}% etc
% \thanks is optional - remove next line if not needed
% \thanks{\emph{Present address:} Insert the address here if needed}%
}                     % Do not remove
%
%\offprints{}          % Insert a name or remove this line
%
\institute{High Energy Physics,  Dept. of Nuclear and Particle Physics, Uppsala University, Box 535, 751 21 Uppsala, Sweden
% \and the second institute address here
}
%
%\date{Received: date / Revised version: date}
% The correct dates will be entered by Springer
\date{}
\abstract{
Isospin asymmetry in $b \rightarrow s \gamma$ transitions provides severe constraints on new physics parameters, due to the small values of the experimental measurements. We exploit here this new observable to study different supersymmetric models. For this purpose, we calculate the isospin symmetry breaking in the MSSM with minimal flavor violation, considering NLO contributions. We study three scenarios for supersymmetry breaking, namely mSUGRA, AMSB and GMSB. In most of the explored regions, we find very restrictive constraints on the allowed parameters for these models.%
 % end of PACS codes
} %end of abstract
\maketitle
\section{Introduction}
\label{intro}
Constraining parameters of new physics, and in particular supersymmetry which is one of the best motivated scenarios beyond the Standard Model, is of utmost importance in preparation for the LHC. Rare $B$ decays have been extensively used for this purpose during the last few years. In fact, Flavor Changing Neutral Currents (FCNC) are forbidden at tree level, and therefore their Standard Model contributions are very small, not allowing much room for new physics contributions. The inclusive branching ratio corresponding to $b \rightarrow s \gamma$ provides restrictive constraints for instance \cite{isidori}. In this context, we are introducing a new observable in $b \rightarrow s \gamma$ transitions, namely isospin asymmetry, which as we will show, provides even more restrictive constraints. \\
\\
The isospin asymmetry in $B \rightarrow K^* \gamma$ decays is defined as:
\begin{equation}
\Delta_{0-}=\frac{\Gamma (\bar B^0\to\bar K^{*0}\gamma ) -\Gamma (B^-\to K^{*-}\gamma )}{\Gamma (\bar B^0\to\bar K^{*0}\gamma )+\Gamma (B^-\to K^{*-}\gamma
)}\;\; . \label{isospinasym}
\end{equation}
This asymmetry has been measured experimentally by Babar \cite{babar} and Belle \cite{belle}:
\begin{eqnarray}
\Delta_{0-}&=& +0.050 \pm 0.045({\rm stat.})\nonumber\\
&&\pm 0.028({\rm syst.})\pm 0.024(R^{+/0})\;\;\; (\mbox{Babar})\; , \label{babar}\\
\Delta_{0+}&=& +0.012 \pm 0.044({\rm stat.})\nonumber\\
&&\pm 0.026({\rm syst.})\;\;\; (\mbox{Belle})\; . \label{belle}
\end{eqnarray}
Calculating the isospin asymmetry, while considering the supersymmetric contributions, and comparing the results with the above experimental data allows us to establish very tight constraints on the SUSY parameters \cite{Ahmady,mahmoudi}.\\
\\
In this study we investigate different supersymmetry breaking scenarios, such as minimal supergravity \newline (mSUGRA) model \cite{msugra1,msugra2}, minimal Anomaly Mediated SUSY Breaking (AMSB) \cite{amsb} and the minimal Gauge Mediated SUSY Breaking (GMSB) \cite{gmsb} scenarios. The method of calculation of the isospin asymmetry, as well as the results for different supersymmetry breaking scenarios are presented in the following sections.
%%%%%%%%%%%
\section{Isospin asymmetry calculation and the SuperIso package}
\noindent Using the QCD factorization and following the method of \cite{kagan}, one can show that the isospin symmetry breaking, $\Delta_{0-}$, can be written as:
\begin{equation}
\Delta_{0-} =\mbox{Re}(b_d-b_u) \;\;,
\end{equation} 
where the coefficients $b_q$ reads:
\begin{equation}
b_q = \frac{12\pi^2 f_B\,Q_q}{\bar m_b\,T_1^{B\to K^*} a_7^c}\left(\frac{f_{K^*}^\perp}{\bar m_b}\,K_1+ \frac{f_{K^*} m_{K^*}}{6\lambda_B m_B}\,K_{2q} \right)\;\;.
\end{equation}
In this formula, the coefficients  $a_7^c$, $K_1$ and $K_{2q}$ can be written in function of the Wilson coefficients $C_i$ at scale $\mu_b$. The other parameters are described in \cite{Ahmady}, where we have detailed the method of the calculation. \\
\\
The calculation of isospin asymmetry has been implemented in a C program, SuperIso \cite{superiso}, which can calculate the Wilson coefficients, the isospin symmetry breaking of $B \to K^* \gamma$, and the inclusive branching ratio of $B \to X_s \gamma$. These calculations can be performed for the mSUGRA, AMSB and GMSB models. 
The SUSY mass spectrum, as well as the couplings and the mixing matrices, can be generated automatically by a call to Softsusy \cite{softsusy} or Isajet \cite{isajet}, or they can be provided by the user in a SUSY Les Houches Accord (SLHA) \cite{slha} file.\\ 
\\
The results presented in the following, for both the inclusive branching ratio and the isospin asymmetry have been calculated using SuperIso v1.0. The SUSY mass spectrum, as well as the couplings and the mixing matrices, are generated using SOFTSUSY 2.0.14 \cite{softsusy} or Isajet 7.75 \cite{isajet}.
%%%%%%%%%%%
\section{Constraints from isospin symmetry breaking}
\label{sec:1}
\begin{figure*}
\includegraphics[width=8.0cm,height=6.7cm]{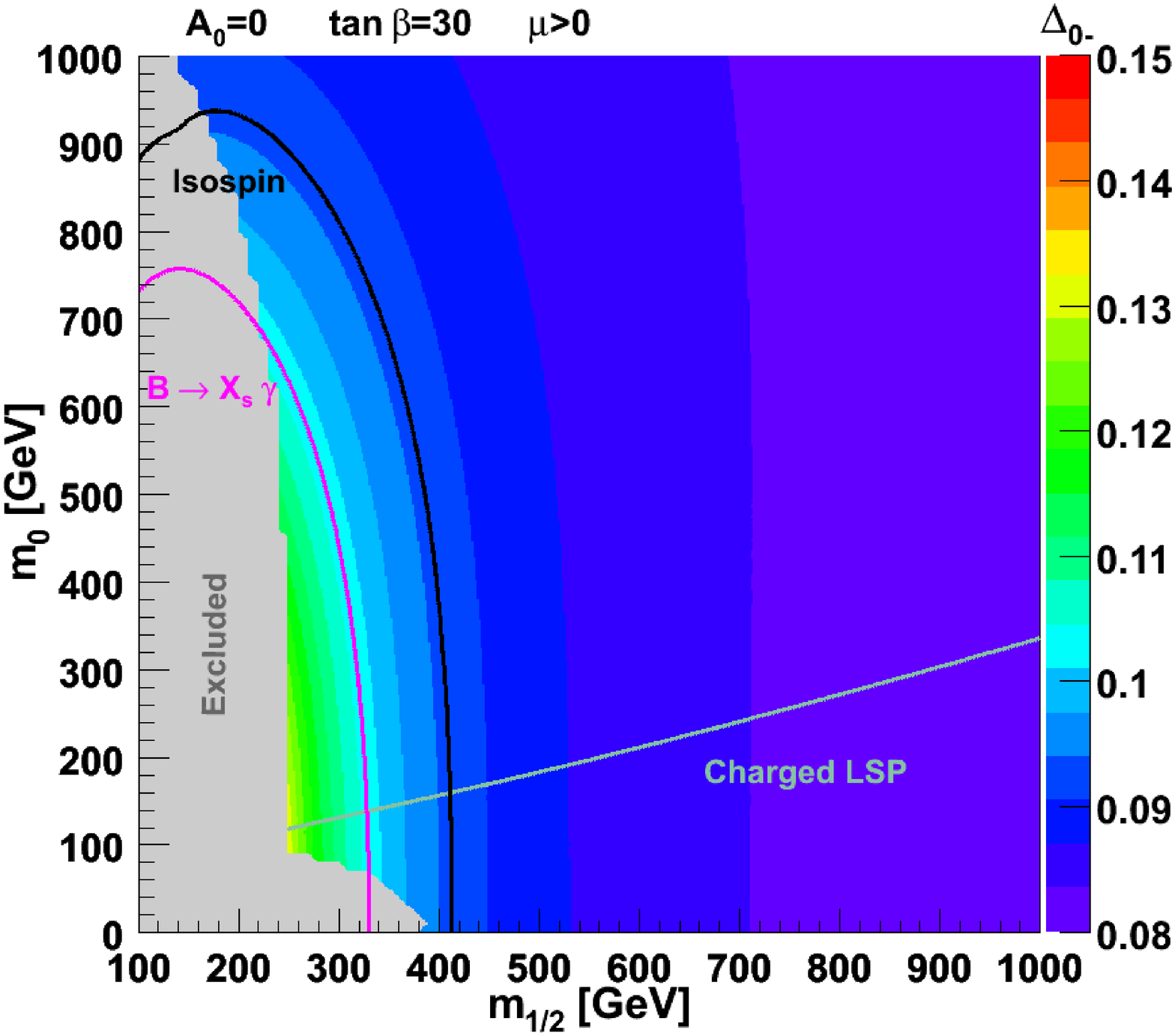}~~~~~\includegraphics[width=8.0cm,height=6.7cm]{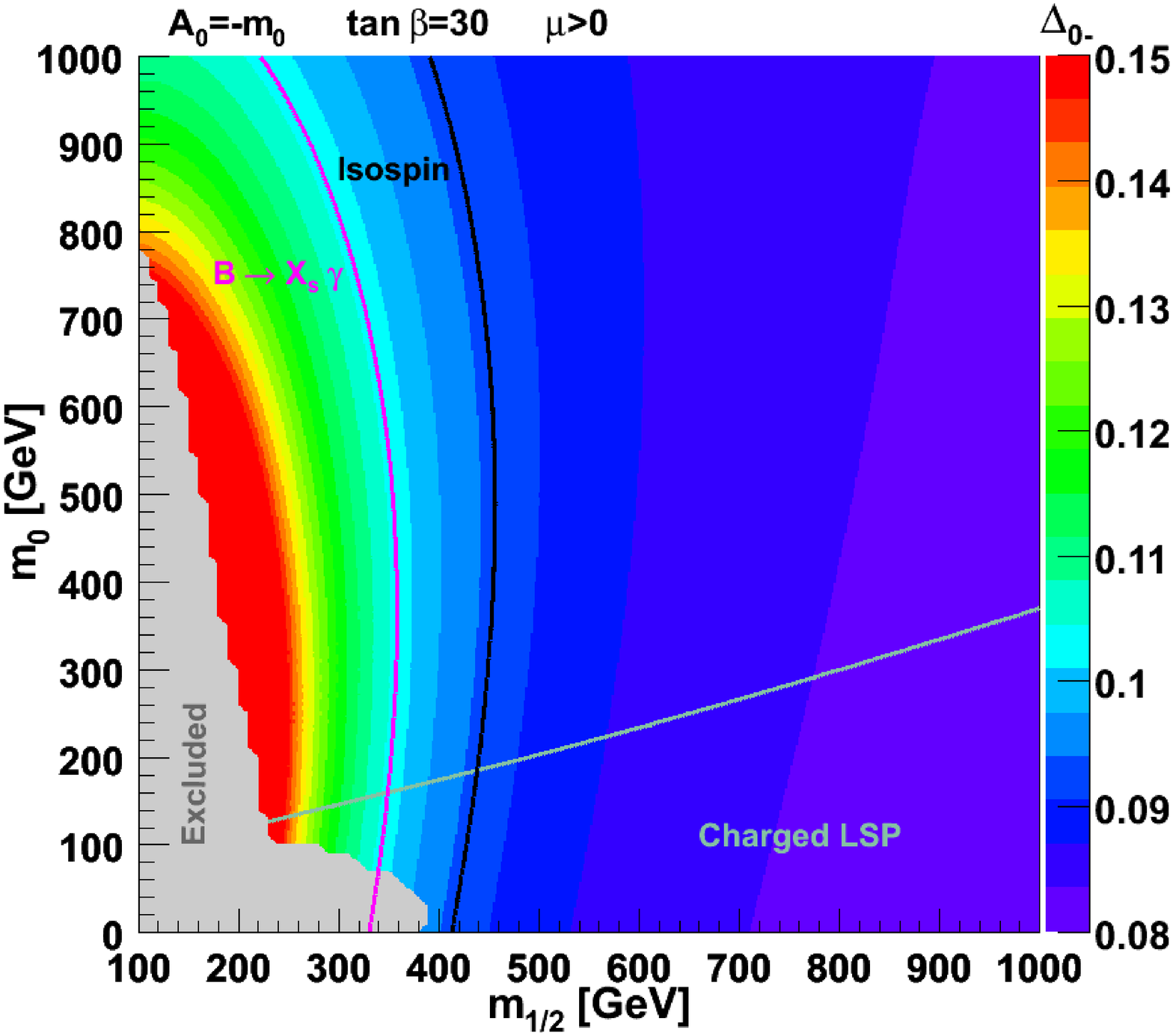}
\caption{Constraints on the mSUGRA parameter plane $(m_{1/2},m_0)$ with $\mu > 0$ for $\tan\beta=30$ and for $A_0=0$ (left) and $A_0=-m_0$ (right). The conventions for the different colors and regions are explained in the text.}
\label{fig:1}       % Give a unique label
\end{figure*}

\begin{figure*}
\includegraphics[width=8.0cm,height=6.7cm]{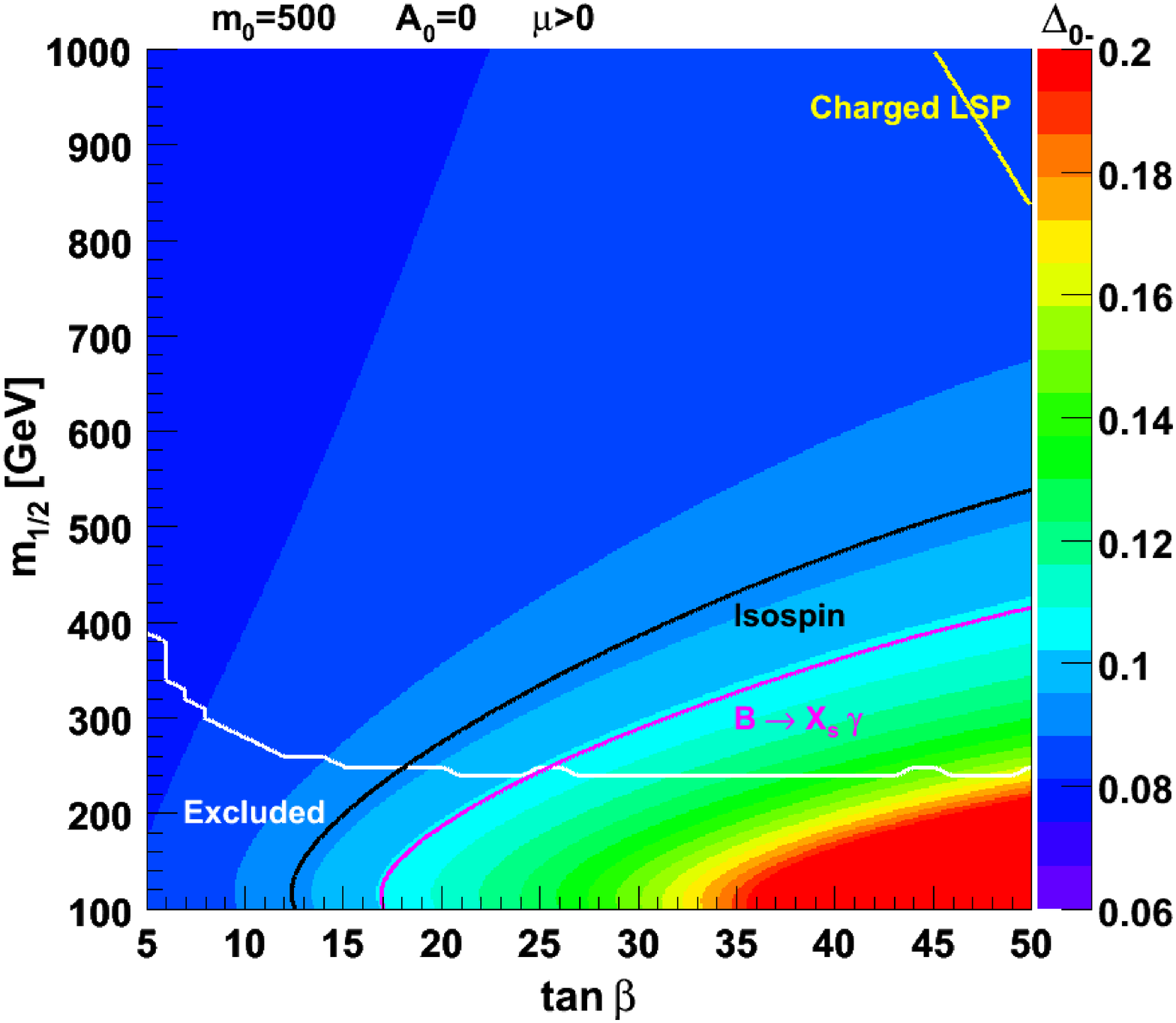}~~~~~\includegraphics[width=8.0cm,height=6.7cm]{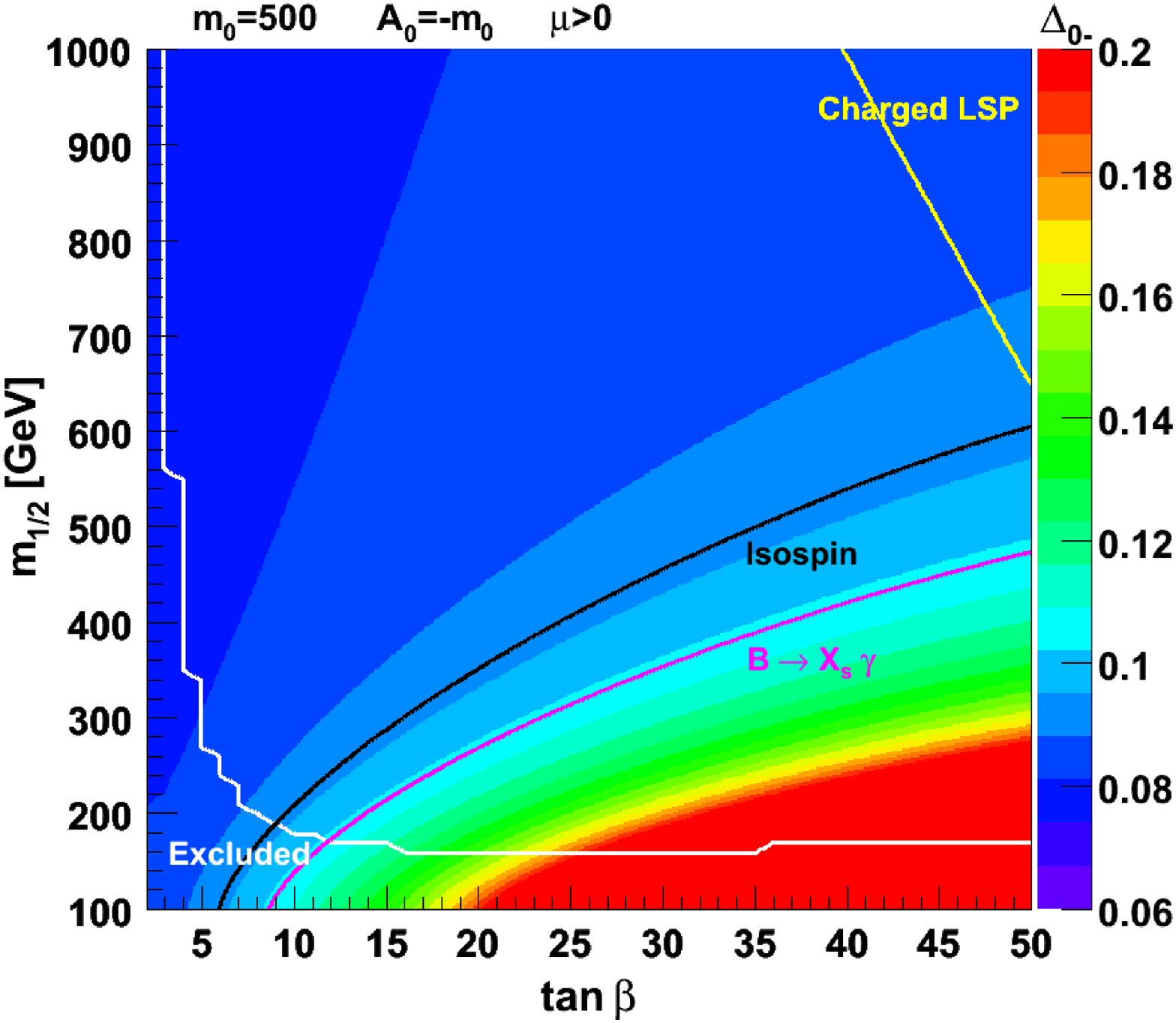}
\caption{Constraints on the mSUGRA parameter plane $(\tan\beta,m_{1/2})$ for $m_0=500$, with $A_0=0$ (left) and $A_0=-m_0$ (right). The definitions of the different regions are explained in the text. Note that the color scale here is different.}
\label{fig:2}       % Give a unique label
\end{figure*}

\begin{figure*}
\includegraphics[width=8.0cm,height=6.7cm]{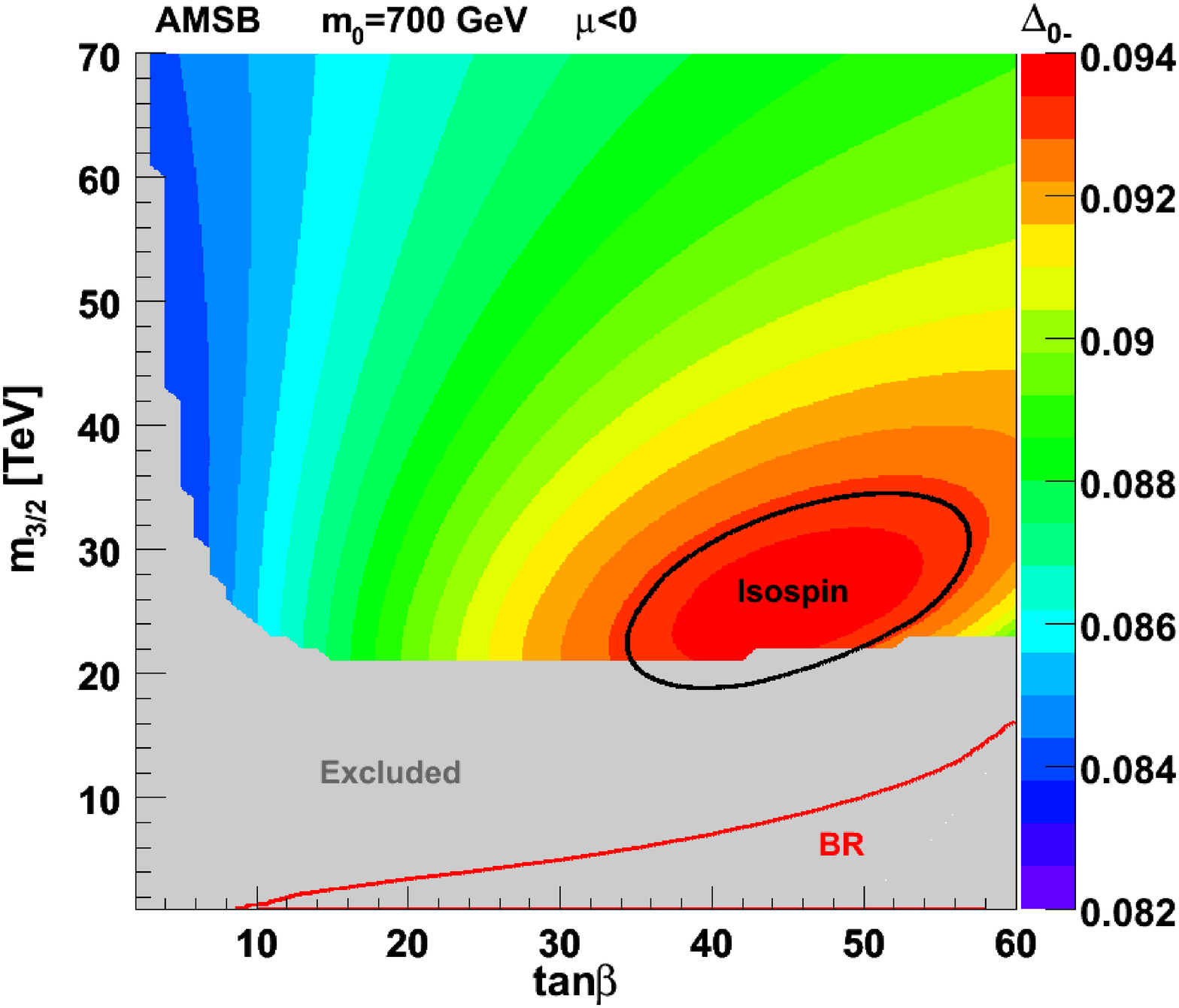}~~~~~\includegraphics[width=8.0cm,height=6.7cm]{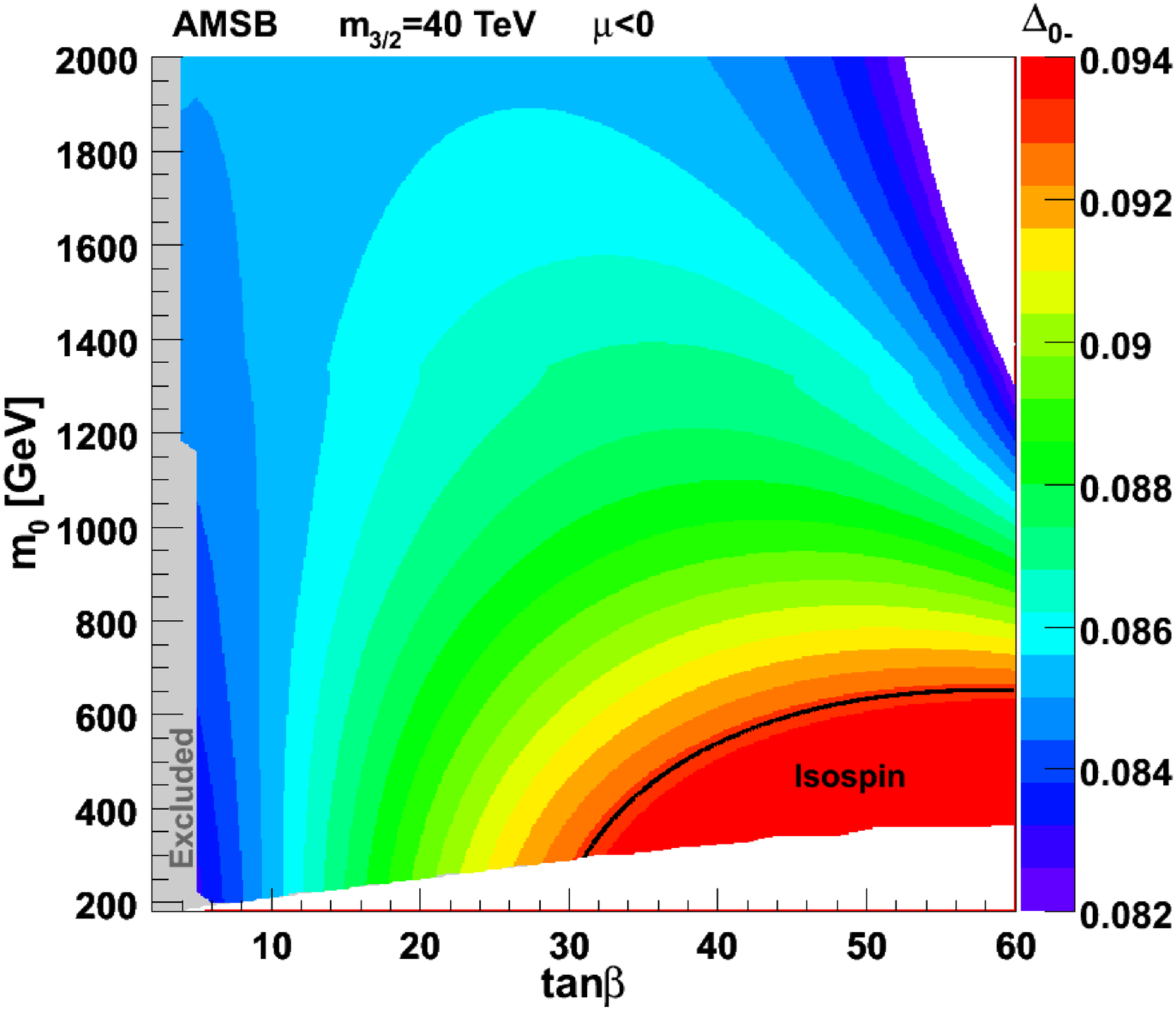}
\caption{Constraints on the AMSB parameter plane $(\tan\beta,m_{3/2})$ for $m_0=700$ GeV to the left and the plane $(\tan\beta,m_0)$ for $m_{3/2}=40$ TeV to the right. In the white regions in the right plot, tachyonic particles can be found. Note that the color scale is different from the previous plots.}
\label{fig:3}       % Give a unique label
\end{figure*}

\begin{figure*}
\includegraphics[width=8.0cm,height=6.8cm]{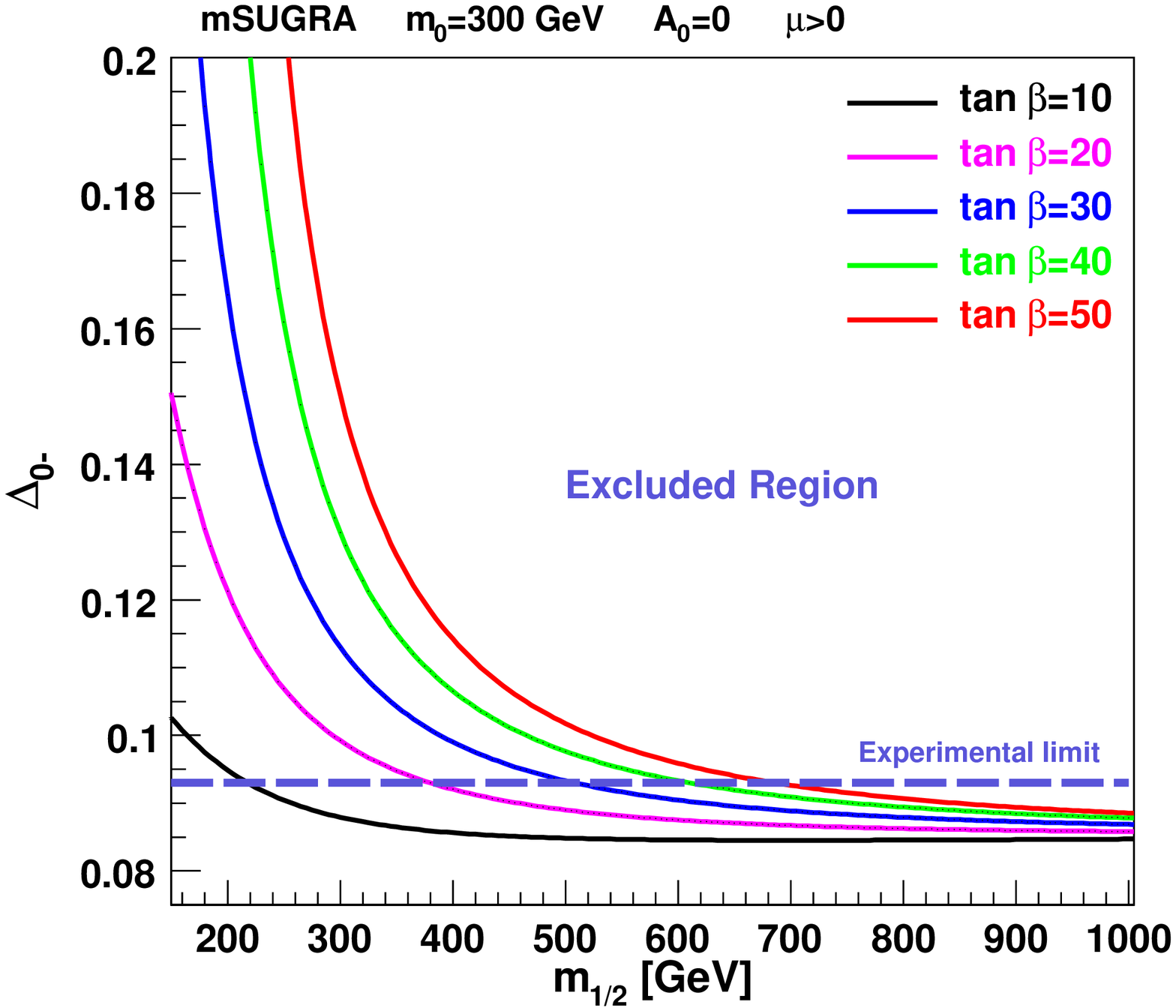}~~~~~\includegraphics[width=8.0cm,height=6.8cm]{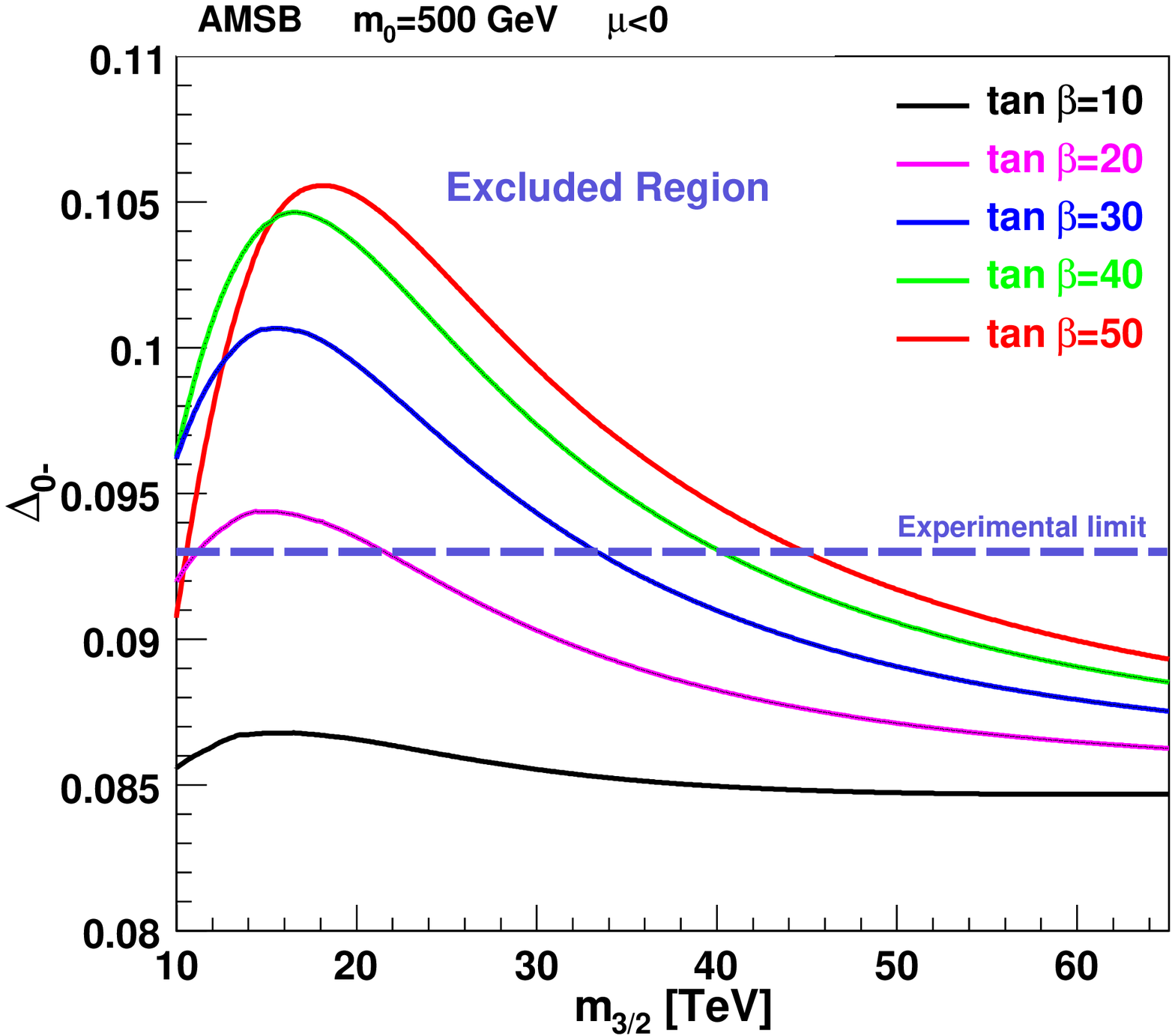}
\caption{Isospin asymmetry in function of $m_{1/2}$ in the mSUGRA parameter space with $m_0=300$ GeV, $A_0=0$ and $\mu >0$ to the left, and in the AMSB parameter space with $m_0=500$ and $\mu < 0$ in function of $m_{3/2}$ to the right. The horizontal dashed line corresponds to the constraint (\ref{isospinconstraint}).}
\label{fig:4}       % Give a unique label
\end{figure*}

\begin{figure*}
\includegraphics[width=8.7cm,height=6.9cm]{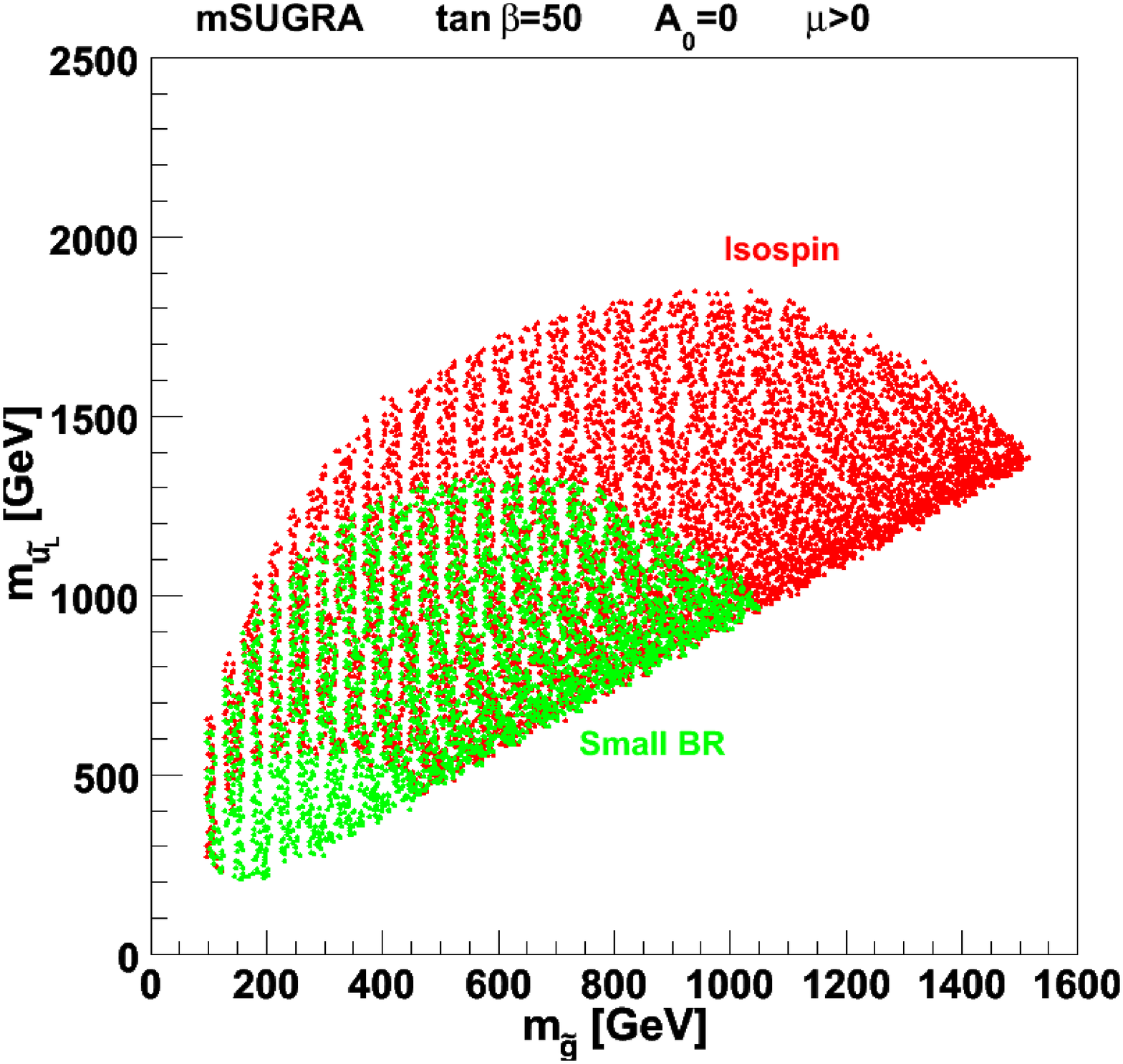}~~~\includegraphics[width=8.5cm,height=6.9cm]{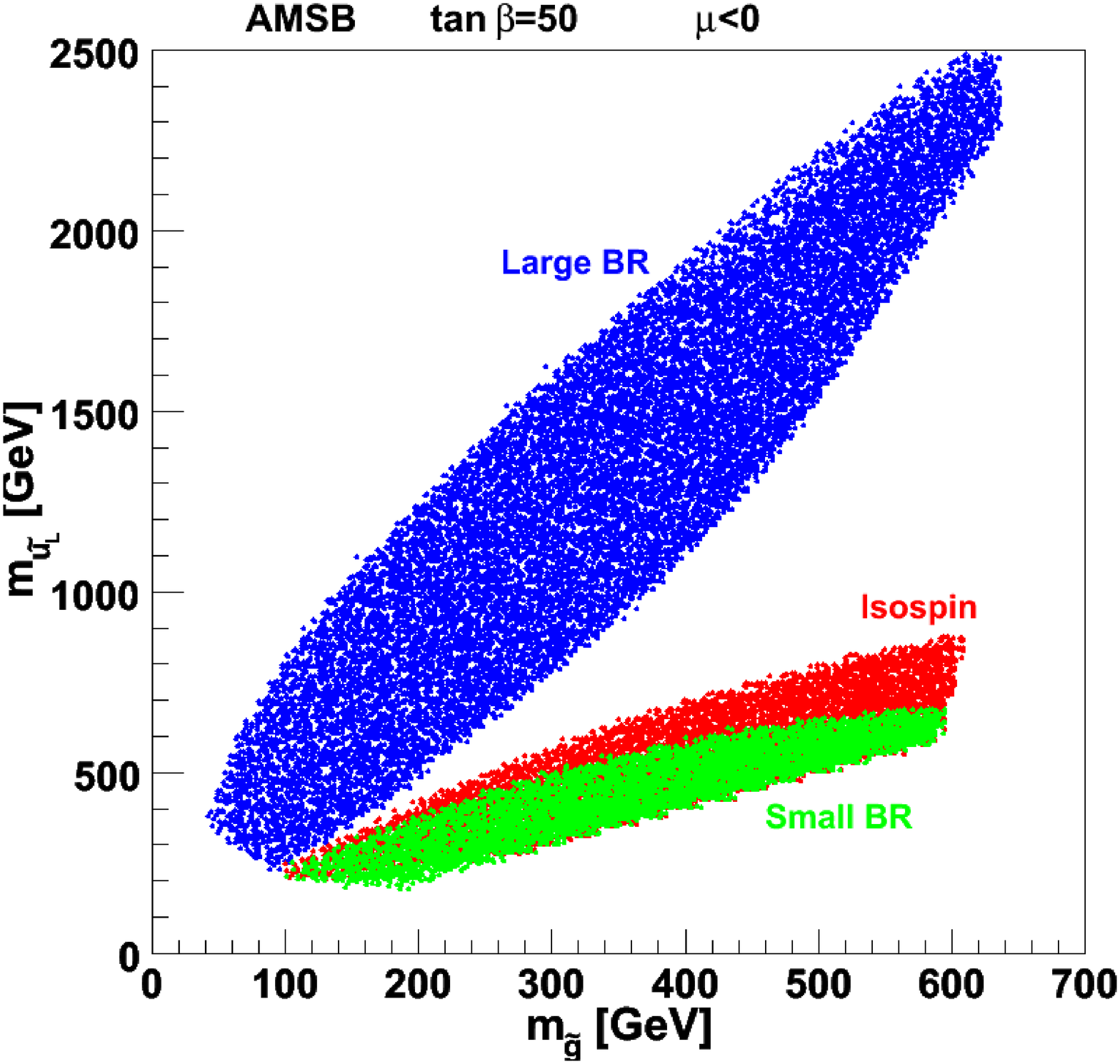}
\caption{($m_{\tilde{g}}$, $m_{\tilde{u}_L}$) plane in the mSUGRA parameter space with $\tan\beta=50$, $A_0=0$ and $\mu >0$ to the left, and in the AMSB parameter space with $\tan\beta=50$ and $\mu<0$ to the right. The red dotted regions are excluded by isospin asymmetry constraints, while the green regions are excluded by the lower bound on the branching ratio and the blue region by the upper bound on the branching ratio.}
\label{fig:5}       % Give a unique label
\end{figure*}

In order to explore the constraints from isospin asymmetry, we scan over parts of the parameter spaces of the supersymmetric models, and for every point we calculate the isospin asymmetry. As a comparison reference, we also calculated the inclusive branching ratio associated to $b \rightarrow s \gamma$.\\
\\
As an example of the results in the mSUGRA parameter space, we investigate the $(m_{1/2},m_0)$ plane for $A_0=-m_0$ and $A_0=0$, with $\tan\beta = 30$. For every parameter space point, the calculated isospin asymmetry is then compared to the combined experimental limits of Babar and Belle. After including the theoretical errors due to the scales and model parameters, the allowed values for isospin asymmetry at $95\%$ confidence level stand at \cite{mahmoudi}:
\begin{equation}
-0.018 < \Delta_{0-} < 0.093 \;\;,\label{isospinconstraint}
\end{equation}
and we use the following limits at 95\% C.L. for the inclusive branching ratio \cite{battaglia}:
\begin{equation}
2.33 \times 10^{-4} < \mathcal{B}(b \to s \gamma) < 4.15 \times 10^{-4} \;\;.
\end{equation}
The results for mSUGRA are shown in Figure~\ref{fig:1}. In this figure, the black contour marked ``Isospin'' corresponds to the region excluded by the isospin breaking constraints, whereas the contour marked ``BR'' corresponds to the region excluded by the inclusive branching ratio constraints. The ``Excluded'' area in the figure corresponds to the case where at least one of the sparticle masses does not satisfy the collider constraints or where the neutral Higgs boson becomes too light \cite{PDG2006}. Finally, the ``Charged LSP'' region is cosmologically disfavored when R-parity is conserved. The various colors represent the changing magnitude of the isospin asymmetry. \\
One can remark here that the isospin asymmetry greatly enlarges the exclusion contours compared to the inclusive branching ratio. Also, since the supersymmetric loop corrections are proportional to the gluino mass and $\tan\beta$, the SUSY contributions can be quite large in the high $\tan\beta$ limit. This enhancement is presented in Figure~\ref{fig:2}. \\
\\
We can now go on to investigate other SUSY breaking mechanisms, such as AMSB and GMSB scenarios.\\
\\
First, we perform scans in the AMSB parameter space $\lbrace m_0$, $m_{3/2}$, $\tan\beta$, $\mathrm{sign}(\mu) \rbrace$.
An investigation of the $(\tan\beta,m_0)$ and $(\tan\beta,m_{3/2})$ planes is presented in Figure~\ref{fig:3}. The conventions in this figure are the same as in the precedent figures. Note that the isospin asymmetry contours, contrary to the cases in the precedent figures, are disconnected from the branching ratio. In fact, while the branching ratio does not provide any interesting limit for the studied parts of the parameter space, the isospin asymmetry provides us with some appreciable limits.\\
\\
%
% For two-column wide figures use
%
In Figure~\ref{fig:4}, we also present 2-dimensional plots in the mSUGRA and AMSB parameter spaces to illustrate the running of isospin asymmetry as a function of $\tan \beta$ in a more readable manner. The horizontal dashed line corresponds to the upper limit of the criterion (\ref{isospinconstraint}). We notice again that the isospin asymmetry is very sensitive to the value of $\tan\beta$.\\
\\
For the GMSB scenario, we explored several regions in the parameter space $\lbrace \Lambda$, $M_{\mathrm{mess}}$, $N_5$, $\tan\beta$, $\mathrm{sign}(\mu) \rbrace$. Unfortunately, the available experimental data do not allow us to obtain any constraints from neither the branching ratio nor the isospin asymmetry.\\
\\
Up to here, we studied only the constraints on the SUSY parameters for different models. To have a better idea of the implication of this new constraint on the mass of supersymmetric particles, we investigate now the limits on the masses of squarks and gluinos since these are relevant for the strong interaction phenomenology.\\
The mSUGRA and AMSB parameter spaces have been investigated for this purpose. The results of the scans for ($m_{\tilde{g}}$, $m_{\tilde{u}_L}$) planes are shown in Figure~\ref{fig:5}. \\
The regions excluded by isospin asymmetry are marked with red dots, while those excluded by the branching ratio are marked with blue and green dots. The figure shows clearly the strong constraints imposed by the isospin asymmetry, in particular for mSUGRA model. In the AMSB parameter space, both the upper and lower bounds of the branching ratio provide restrictive constraints, while only the upper bound on isospin asymmetry is used.
%%%%%%%%%%%
\section{Summary}
\label{sec:6} 
We have shown that the isospin asymmetry is a valuable observable for constraining the MSSM parameter space, especially for the mSUGRA and AMSB mechanisms. The SuperIso package provides the possibility to explore the supersymmetry parameter space for different scenarios. New models, such as the Non Universal Higgs Model (NUHM), can also be studied with SuperIso provided all the corresponding masses and couplings are written in a SLHA file. A deeper analysis of the constraints provided by isospin asymmetry can be found in \cite{mahmoudi}. We refer the reader to this article for more details.
%%%%%%%%%%%
%\subsubsection{Subsubsection title} 

%        Comments: Submitted for the SUSY07 proceedings, 4 pages, LaTeX, 3 eps figures.
%
% % For tables use
% \begin{table}
% \caption{Please write your table caption here}
% \label{tab:1}       % Give a unique label
% % For LaTeX tables use
% \begin{tabular}{lll}
% \hline\noalign{\smallskip}
% first & second & third   \\
% \noalign{\smallskip}\hline\noalign{\smallskip}
% number & number & number \\
% number & number & number \\
% \noalign{\smallskip}\hline
% \end{tabular}
% % Or use
% \vspace*{1cm}  % with the correct table height
% \end{table}

% For one-column wide figures use
% \begin{figure}
% % Use the relevant command for your figure-insertion program
% % to insert the figure file.
% % For example, with the option graphicx use
% \includegraphics[width=0.45\textwidth,height=0.15\textwidth,angle=0]{susy07.eps}
% \caption{Please write your figure caption here}
% \label{fig:1}       % Give a unique label
% \end{figure}
%
% For two-column wide figures use
% \begin{figure*}
% % Use the relevant command for your figure-insertion program
% % to insert the figure file. See example above.
% % If not, use
% \includegraphics[width=1.\textwidth,height=0.34\textwidth,angle=0]{schloss_mHwr_templ.eps}
% %\hfill
% %\includegraphics[width=0.45\textwidth,height=0.14\textwidth,angle=0]{susy07.eps}
% \caption{Please write your figure caption here}
% \label{fig:2}       % Give a unique label
% \end{figure*}

%
% BibTeX users please use
% \bibliographystyle{}
% \bibliography{}
%
% Non-BibTeX users please use

\end{document}